# High mobility conducting channel at semi-insulating GaAs - metal oxide interfaces.

G. Kopnov and A. Gerber

Raymond and Beverly Sackler Faculty of Exact Sciences,

School of Physics and Astronomy, Tel Aviv University, Tel Aviv 6997801, Israel

Absence of an efficient technology of GaAs passivation limits the use of III-V semiconductors in modern electronics. The effect reported here can possibly lead to a solution of this long standing problem. We found that an electrically conducting interfacial channel is formed when insulating metal oxide dielectrics are deposited on untreated semi-insulating GaAs wafers by reactive RF sputtering in argon/oxygen plasma. The conducting channel is n-type with the surface charge density of $10^7$ to $10^{10}$ cm$^{-2}$ and Hall mobility as high as 6000 cm$^2$/Vsec, depending on the RF plasma excitation power and oxygen content during the deposition. The conducting channel is formed by depositing any of the tested metal oxide dielectrics: MgO, SiO$_2$, Al$_2$O$_3$ and HfO$_2$.

Corresponding author: A. Gerber, gerber@tauex.tau.ac.il

# Introduction.

Electron mobility in bulk III-V semiconductors is about an order of magnitude higher than in silicon, which promises much faster nanometer-scale electronics in GaAs-based devices [1]. However, fabrication of metal oxide - semiconductor field effect transistors (MOSFETs), the central element of the modern electronics, with III-V semiconductors is hampered for years by the unresolved problem of the semiconductor to metal-oxide interface. Exposure of the III–V surface to oxygen results in appearance of the interfacial states that reduce the charge mobility and cause the 'Fermi-level pinning', which is an inability to modulate the electrostatic potential inside the semiconductor [2]. The defects are attributed to a number of native surface species including Ga and As oxides and sub-oxides, elemental As, As–As dimers and Ga dangling bonds [3]. Multiple methods of GaAs surface passivation were tried, including formation of native oxides by thermal, anodic and plasma oxidations, and deposition of various insulators by thermal, plasma, chemical vapor deposition and other more exotic techniques [4]. All these efforts had limited success. On this background one should note two discoveries important for preserving optimism and stimulating the ongoing research. The first was the discovery that a stable oxide $Ga_2O_3$ deposited *in situ* on GaAs yields an interface free of the Fermi-level pinning with the quality close to that of the AlGaAs/GaAs interface [5]. Unfortunately, $Ga_2O_3$ by itself is not suitable to serve as a high permeability insulator in the MOSFET stack because of high oxide bulk trap densities and excessive leakage current [6]. The second advance was the discovery of the interface self-cleaning effect that takes place when oxides like $Al_2O_3$ are deposited on GaAs by atomic layer deposition (ALD) [7, 8]. The native GaAs surface oxides are largely eliminated at the very early stages of ALD, and subsequent exposure to the ALD oxidant does not regrow the III–V oxides.

Here, we report on discovering another approach to fabrication of high mobility interfaces between GaAs and metal oxide dielectrics. We found that exposure of untreated semi-insulating GaAs wafers to $Ar/O_2$ plasma during the reactive RF sputtering of metal oxides leads to self-cleaning and formation of the n-type conducting interfacial channel with mobility close to that found in GaAs/AlGaAs heterostructures.

**Experimental.**

Metal oxide dielectrics, including MgO, SiO$_2$, Al$_2$O$_3$ and HfO$_2$, were RF sputtered on 5mm x 5mm slices of commercial SI-GaAs [001] wafers at room temperature. 500 µm thick compensated wafers, usually used as insulating substrates for film deposition, were supplied by different manufacturers (CMK and XTAL). Their room temperature resistance in dark exceeded 20 GΩ, but decreased to $10^9$ -$10^{10}$ Ω in the ambient laboratory illumination due to generation of the light excited carriers. No special chemical or thermal treatment of the wafers was done but a standard procedure of the Ohmic contacts fabrication. Sputtering was performed at 4 to 5 mtorr pressure of argon with the added flow of air or 99.999% oxygen. The base chamber pressure prior to deposition was about $1\times10^{-7}$ torr. Hygroscopic targets, such as MgO, previously exposed to an ambient air were cleaned by pre-sputtering for 20 minutes at 5 W/cm$^2$ RF power density to remove the contaminated surface layer. The RF power density delivered to the targets varied between 2.5 W/cm$^2$ and 12 W/cm$^2$. Only a limited number of samples were fabricated at 12 W/cm$^2$ due to an excessive heat load. The deposition rate varied between 0.1 nm/min to 0.3 nm/min, subject to the applied power, target material and oxygen content in the chamber. Most of the samples were deposited in the custom built RF sputtering system using 2-inch targets (ACI alloys Inc.) located 20 cm from the substrate. A number of control samples were deposited in the commercial Vinci Technologies magnetron sputtering installation. All samples fabricated in both systems demonstrated similar properties. Two types of electric contacts were used: (i)Ni/Ge/Au Ohmic contacts grown on GaAs substrate prior to depositing the metal oxides films; (ii) alternatively, Al/Si wires were bonded to the film surface after the deposition. Samples with both types of contacts exhibited linear I-V characteristics and revealed the same results. Resistance and Hall effect measurements were done using the Van der Pauw four contacts protocol. Four probe and two-probe measurements using the Keithley electrometer were applied for the high resistance samples in the GΩ range. The results of the two probe measurements were consistent with the four probe ones. Structural analyzes were done by X- ray diffraction, high-resolution transmission electron microscopy, cross-sectional TEM, TOF-SIMS and XPS.

# Results and discussion.

Figure 1 presents the resistance of the initially insulating GaAs wafer samples measured *in situ* during the RF sputtering deposition of MgO in Ar/O$_2$ atmosphere at three different partial pressures of argon and oxygen. Electric contacts were attached to the substrate prior to the deposition. The actual four-probe resistance measurement in the Van der Paw configuration was performed at given time intervals with switched-off RF power. Deposition was resumed after each measurement. Resistance in dark of the GaAs substrate prior to the deposition exceeded 20 GΩ and remained immeasurable when MgO was sputtered in a pure Ar atmosphere. The sample remained insulating, as would be expected when an insulating metal oxide is deposited on top of an insulating substrate. When oxygen was added to Ar, the resistance started dropping immediately with the beginning of sputtering. The resistance stopped decreasing after about 30 minutes of deposition when an average thickness of the grown MgO layer exceeded 2nm. No significant changes were observed when more MgO was deposited. Since the bulk of GaAs wafer was insulating and the final resistance did not depend on the thickness of the deposited film, we attribute the conductance to the interfacial layer formed between GaAs and MgO. Similar depositions performed in the mixed argon/nitrogen atmosphere produced no measurable conductance. Reactive sputtering of MgO in Ar/O$_2$ atmosphere on glass, quartz and SiN substrates resulted in the insulating material. Thus, GaAs is the only one of the tested insulating substrates that became conducting during the reactive sputtering of a metal oxide dielectric. We shall denote the samples fabricated by sputtering MgO in presence of oxygen as GaAs(O)/MgO.

Conductance of the interface depends on the deposition conditions including the oxygen concentration, RF power and the substrate used, but also on illumination and time elapsed after the fabrication. Fig.2 presents the resistance of multiple GaAs(O)/MgO samples deposited at different power densities on different wafers as a function of oxygen concentration in the plasma. The measurements were done shortly after the fabrication. The onset of conductance can be traced when as little as 2% of oxygen was added to argon at 12 W/cm$^2$ power density. The resistance decreases gradually with increasing oxygen content up to about 50% where the data become scattered. Notably, the resistance can be as low as $10^5$ Ω/□ in samples fabricated in the oxygen rich plasma.

Density and mobility of the charge carriers were extracted from the Hall effect measurements. Figure 3 illustrates the qualitative difference between the field dependent Hall resistance measured in the sample of untreated GaAs (open circles) and the sample of GaAs(O)/MgO produced on the same wafer (solid circles). The measurements were done in the illuminated laboratory conditions when the zero field resistance of the GaAs sample was 1.3 GΩ. Its Hall resistance $R_{xy}$ is not linear and not monotonic in field, which together with the parabolic in field magnetoresistance is typical for the two-band conductance in the semi-insulating GaAs [9]. Resistance of the GaAs(O)/MgO sample is about 10 MΩ. Its Hall resistance is a linear function of field with a negative Hall effect coefficient $R_H = \frac{dR_{xy}}{dB}$. Linearity of the field dependent Hall resistance indicates the dominance of a single negative charge carrier. The planar charge carrier density $n$ and the Hall mobility $\mu$ were calculated from the resistance and the Hall coefficient as:

$$n = -(R_H e)^{-1}$$

and:

$$\mu = (enR_\square)^{-1} = \frac{R_H}{R_\square}$$

where $R_\square$ is the planar resistance (per square).

Fig. 4 presents the conductance $\sigma = R_\square^{-1}$ of multiple GaAs(O)/MgO samples deposited at different conditions as a function of their carrier density. The carrier density varies between $10^7$ cm$^{-2}$ to $10^{10}$ cm$^{-2}$, while the value of $10^7$ cm$^{-2}$ corresponds to the effective density of the GaAs wafer itself (see Fig.3). Thus, the reactive sputtering produces additional carriers in the conduction band increasing their density by up to 3 orders of magnitude to $10^{10}$ cm$^{-2}$. The conductance follows the carrier density and increases over three orders of magnitude between $10^{-8}$ (Ω/□)$^{-1}$ up to $10^{-5}$ (Ω/□)$^{-1}$ in the sample with the highest carrier density $10^{10}$ cm$^{-2}$. The entire data follow a clear Drude-type correlation between the conductance and the charge carrier density: $\sigma \propto n$.

Mobility as a function of the charge carrier density is shown in Fig.5. Despite a relatively large dispersion of data, mobility of *all* samples produced by the reactive sputtering of MgO on different GaAs wafers at different powers, oxygen concentrations and the overall plasma pressures falls between 1000 cm$^2$/Vsec to beyond 6000 cm$^2$/Vsec. The highest mobility was found systematically in samples fabricated at the highest power density used: 12 W/cm$^2$, marked by the solid symbols in the figure. The effective planar mobility of the illuminated GaAs substrate estimated from the data in Fig.3 is about 10$^2$ cm$^2$/Vsec. Thus, the mobility of the interface is enhanced by at least an order of magnitude and exceeds 6000 cm$^2$/Vsec in at least two samples. This mobility is close to the highest found in GaAs crystals at room temperature and equals to the one obtained in high mobility GaAs/AlGaAs hetero-interfaces with typical charge density of order 10$^{12}$ - 10$^{13}$ cm$^{-2}$ [10]. It exceeds by far the highest values of drift mobility reported in the passivated III-V semiconductors: 500 cm$^2$/Vs in the n-doped GaAs channel with atomic-layer deposited Al$_2$O$_3$ [11], 885 cm$^2$/Vsec in GaAs passivated by a thin amorphous Si cap [12] and the peak value electron mobility of 1190 cm$^2$/Vsec reported in the ALD fabricated InAs MOSFET with ZrO$_2$ dielectric [13].

Transformation of the GaAs interface occurs gradually during the sputtering process. Development of mobility as a function of the sputtering time, together with the respective reduction of resistance is shown in Fig.6. This series of samples were fabricated at 4.9 W/cm$^2$ power density at 1:1 O$_2$/Ar ratio on substrates cut from the same wafer. Resistance and the Hall coefficient were measured outside the sputtering chamber. The mobility (solid circles) increases gradually with the deposition time and saturates when MgO covers the substrate by about 2 nm thick layer. The resistance (open circles) decreases and saturates respectively.

The effect of the RF power on mobility and the carrier density is illustrated in Fig.7. All samples of the series were deposited at the same 1:1 Ar/O$_2$ atmosphere on pieces cut from the same segment of the GaAs wafer. Both the carrier density and the mobility increase with the excitation power and, notably, both seem to follow the same functional dependence of the power.

Resistance of GaAs(O)/MgO samples kept in darkness increases with time and exceeds 20GΩ after a few weeks. Fig.8 illustrates the carrier density and mobility of a typical sample as a function of time elapsed since the fabrication. The aging occurs due to reduction in the charge carrier density. Notably, the mobility does not deteriorate with time and remains constant, which indicates an effective long term passivation of the interface.

Formation of the conducting channel during the reactive sputtering of dielectrics on GaAs is a general property. Fig.9 presents the resistance measured *in situ* during the reactive sputtering of four principal metal oxide dielectrics: MgO, SiO$_2$, Al$_2$O$_3$ and HfO$_2$ on the initially insulating GaAs. Development of a conducting channel is common to all of them: the resistance starts dropping immediately with exposing the wafer to the incoming flux and saturates when the deposited dielectric covers the substrate by a continuous layer.

At the moment we possess very little additional information to track the origins of the effect. The interface conduction can develop due to addition or modification of the defect states distribution and formation of the defect level conduction band. Respectively, the carrier density can increase by thermal excitation of electrons via the gap levels to the conduction band. Capacitance – voltage or conductance – voltage measurements, generally used to extract the density of states at metal oxide – semiconductor interfaces [14, 15], require the tunnel resistance of the metal oxide insulator layer to be much higher than that of the semiconductor. Fabrication of such insulator is challenging in our case with an insulating GaAs substrate and $10^6 - 10^7$ Ω/□ conducting interface channel. Therefore, we are unable to support our results by the density of states data. A number of characterization tools were applied in attempt to uncover the compositional or structural uniqueness of GaAs(O)/metal-oxide samples. XPS and TOF-SIMS tests did not reveal any measurable difference between the samples sputtered with and without oxygen. The cross-sectional high resolution TEM analysis of one of the GaAs(O)/SiO$_2$ samples indicated the formation of Ga$_2$O$_3$ interface layer between GaAs and SiO$_2$, however this observation was not confirmed in other samples. Thus, more high quality characterization studies are required.

By analogy with the ALD process that removes the native GaAs surface oxides and stabilizes the pinning free Ga$_2$O$_3$ one, we recall the process of resputtering [16 - 18] that can occur during reactive sputtering. Negative oxygen ions are created close to the magnetron target when the latter is an oxide or the sputtering chamber contains oxygen. These ions are accelerated towards the cathode ring and can bombard the substrate and the growing film with an energy of several hundred eV, while part of the ions lose the attached electron in flight and reach the substrate as energetic neutral oxygen atoms. Resputtering results in a decreased film thickness or even a total suppression of film growth and etching of the substrate. We found clear indications of the resputtering during our sample deposition process, for example, the deposition rate of MgO is reduced by a factor of 2.5 when the working gas mixture contains 1:1 Ar:$O_2$ ratio. One can suggest that bombardment of GaAs surface by energetic oxygen ions, that starts immediately with the exposure of GaAs to the oxygenated plasma, and the simultaneous coverage of the etched surface by the protective dielectric layer can be the mechanism of the self-cleaning and passivation. This scenario is not supported by experimental evidence so far.

Formation of a conducting interface channel at the semi-insulating GaAs – metal oxide interface might be relevant to yet other unresolved problems associated with the III-V electronics, e.g. the side wall leakage current in photodiodes, LEDs and solar cells grown on pristine GaAs. Losses in brightness and efficiency are known to increase with the perimeter to area ratio, which is attributed to the perimeter recombination velocity [19 – 21]. On the other hand, the conducting interface formed between GaAs and the capping metal oxide layer surrounding the device maze can serve the hidden peripheral leakage channel.

**Summary.**

To summarize, two interrelated processes seem to occur during a reactive sputtering of metal oxide dielectrics on semi-insulating GaAs. The first is a quasi-persistent n-type doping of the interface increasing the planar charge carrier density by up to three orders of magnitude. The charge density increases with illumination and decreases slowly in darkness over the weeks-long aging period. The second is the self-cleaning of GaAs surface from the native oxide defects, manifested by an almost ideal mobility of the created conducting channel. The mobility is permanent and independent on the charge density of the aging samples. The conducting channel is formed by sputtering any of the tested metal oxide dielectrics: MgO, $SiO_2$, $Al_2O_3$ and $HfO_2$, therefore the treatment of the interface depends mainly on the process itself and not on the composition of the deposited material. We suggest that the self-cleaning of GaAs might be the result of surface etching by energetic oxygen ions and the simultaneous coverage of the etched surface by a protective dielectric layer. One can expect that the same self-cleaning mechanism can be applied for passivating doped GaAs as well, thus allowing fabrication of high mobility GaAs - metal oxide heterostructures.

## Acknowledgement.

The work was supported by the Israel Science Foundation grant No. 992/17.

**References.**

**Figure captions.**

Fig.1. Development of the conducting channel during reactive sputtering of MgO on semi-insulating GaAs. Resistance of the initially insulating GaAs wafer samples as a function of the deposition time measured *in situ* during the RF sputtering of MgO in Ar/O$_2$ atmosphere. The data were taken during three deposition sequences at 7.4 W/cm$^2$ RF power at argon/oxygen partial pressures of 5/0.25 (open triangles), 5/1.2 (open circles) and 4/4 mtorr (solid circles). The four-probe resistance measurement in the Van der Paw configuration was performed with switched-off RF power. Deposition was resumed after each measurement.

Fig.2. Resistance of multiple GaAs(O)/MgO samples as a function of oxygen concentration in the plasma. The samples were deposited at different power densities on different wafers. The measurements were done outside the sputtering chamber shortly after the deposition.

Fig.3. Hall resistance $R_{xy}$ of the GaAs sample (open circles) cut from the untreated wafer and the GaAs(O)/MgO sample deposited on the same wafer (solid circles) as a function of magnetic field applied normal to the plane. The measurements were done in the illuminated laboratory conditions. The zero field resistance of the GaAs and GaAs(O)/MgO samples were 1.3 GΩ and 10 MΩ respectively.

Fig.4. Conductance $\sigma = R_\square^{-1}$ of multiple GaAs(O)/MgO samples deposited at different conditions as a function of their planar carrier density. Samples deposited at 12.3 W/cm$^2$ power are marked by solid circles.

Fig.5. Hall mobility of multiple GaAs(O)/MgO samples as a function of their planar carrier density. All samples, regardless the details of their fabrication, exhibit mobility exceeding 1000 cm$^2$/Vsec.

Fig.6. Resistance (left vertical axis, open circles) and mobility (right vertical axis, solid circles) of a series of GaAs(O)/MgO samples as a function of the exposure time to the reactive MgO sputtering at $P_{RF} = 7.4$ W/cm$^2$ and 1:1 Ar/O$_2$ ratio.

Fig.7. Planar carrier density (left vertical axis, open triangles) and mobility (right vertical axis, solid circles) of a series of GaAs(O)/MgO samples as a function of the RF sputtering power. The plasma was composed of 1:1 Ar/O$_2$ ratio.

Fig.8. Charge carrier density (left vertical axis) and mobility (right vertical axis) of the GaAs(O)/MgO sample as a function of time elapsed since the fabrication. The measurements were done in dark after removal from the deposition chamber and exposing the sample to the laboratory illumination. Gradual increase of resistance is due to reduction of the charge carriers, while the mobility remains constant.

Fig.9. Development of the conducting channel during reactive sputtering of any of the tested metal oxide dielectrics. Resistance of the initially insulating GaAs samples as a function of the deposition time measured *in situ* during the RF sputtering of of HfO$_2$, MgO, SiO$_2$ and Al$_2$O$_3$ in 1:1 Ar/O$_2$ atmosphere. $P_{RF} = 7.4$ W/cm$^2$.

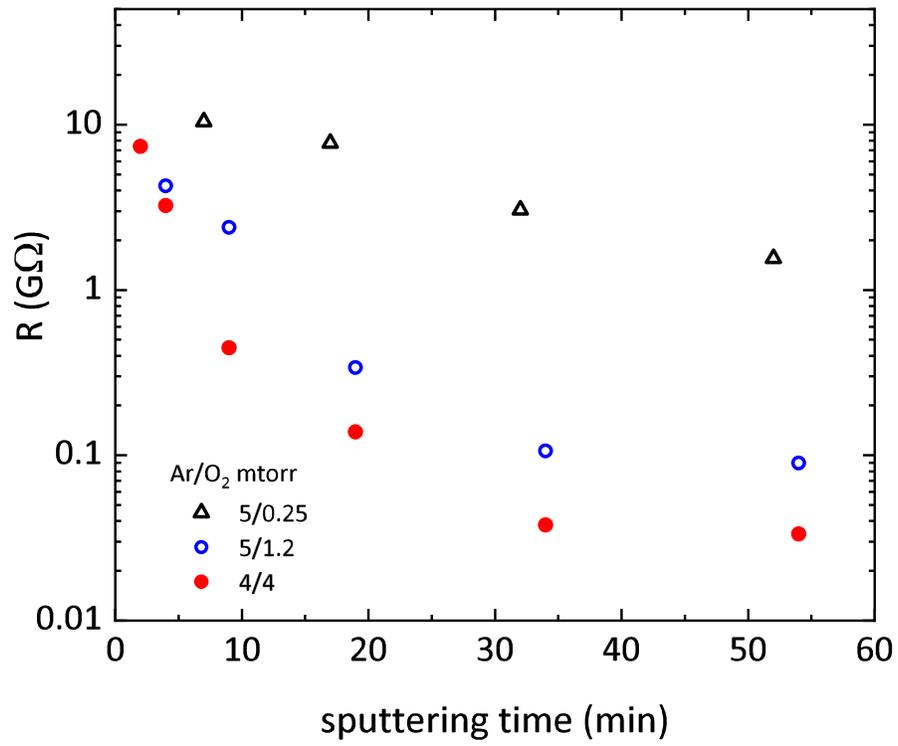

Fig.1

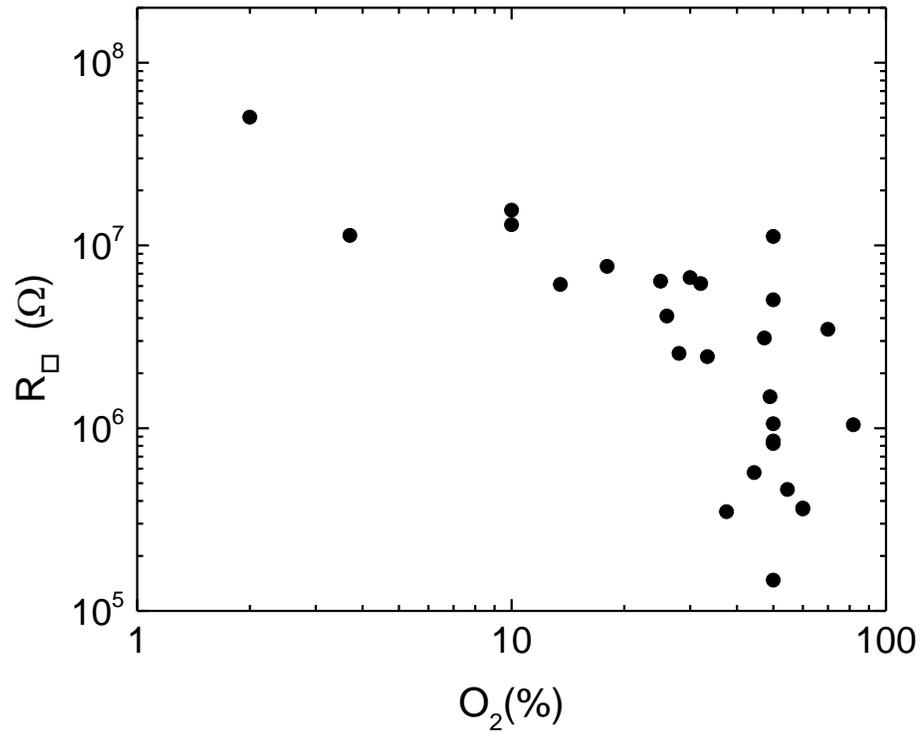

Fig.2

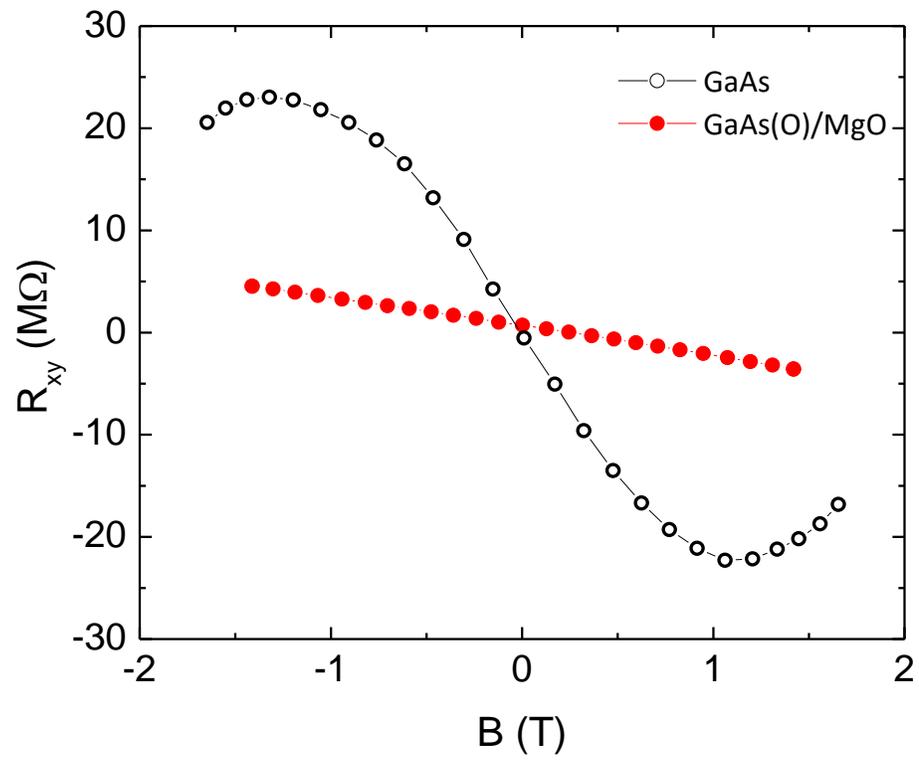

Fig. 3

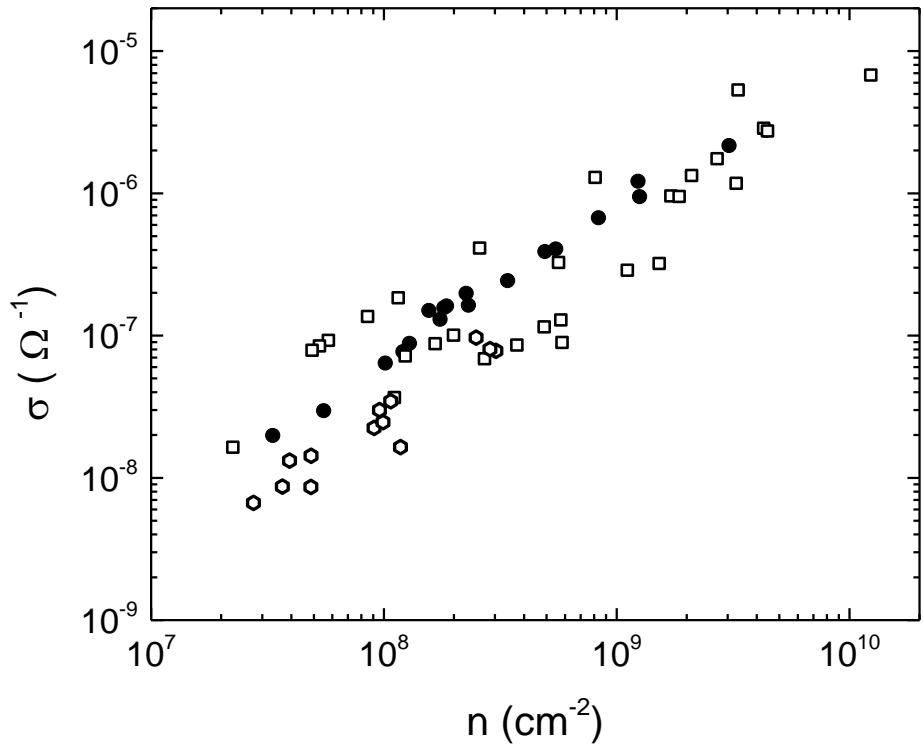

Fig. 4

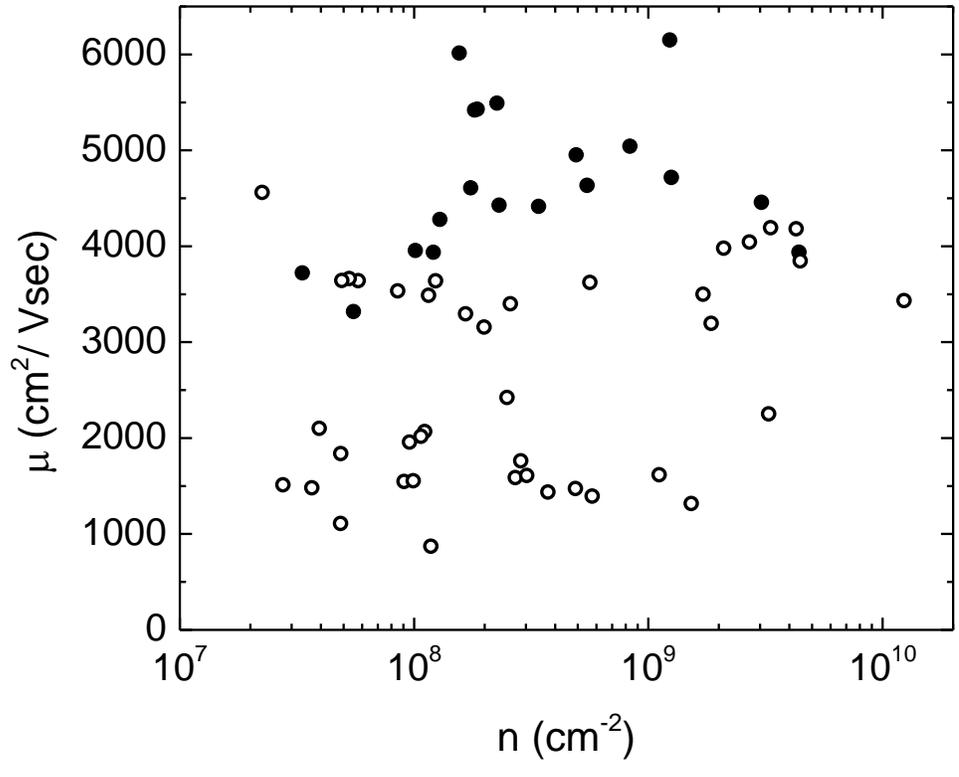

Fig. 5

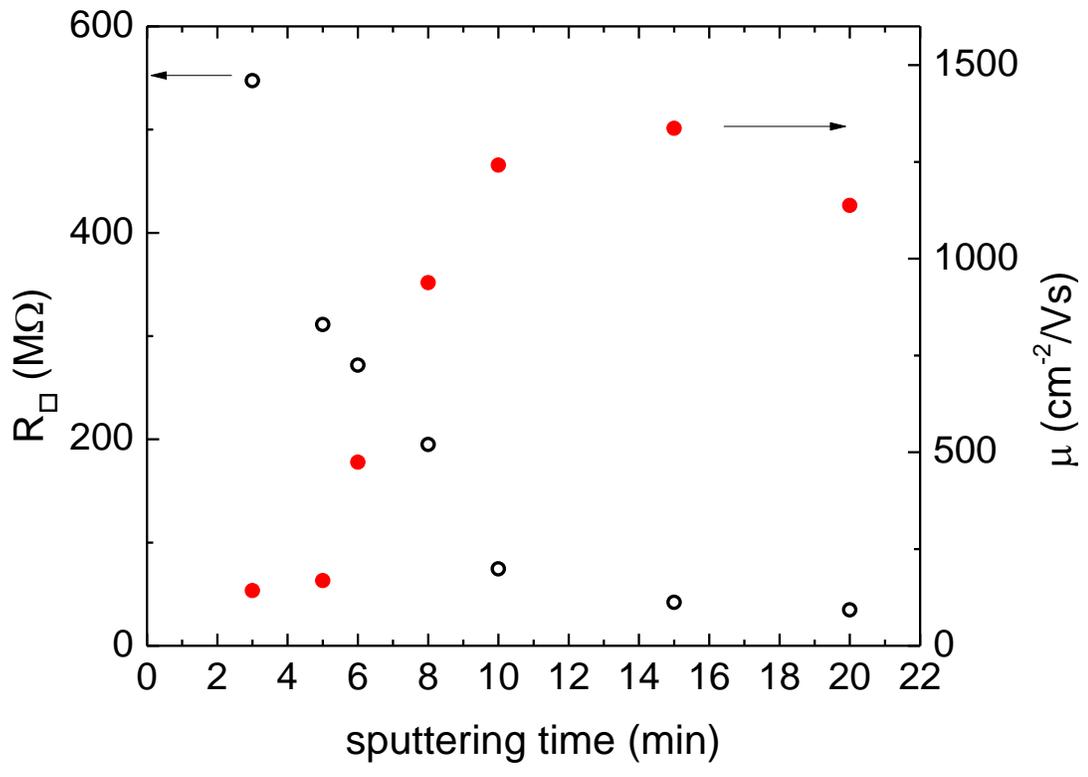

Fig. 6.

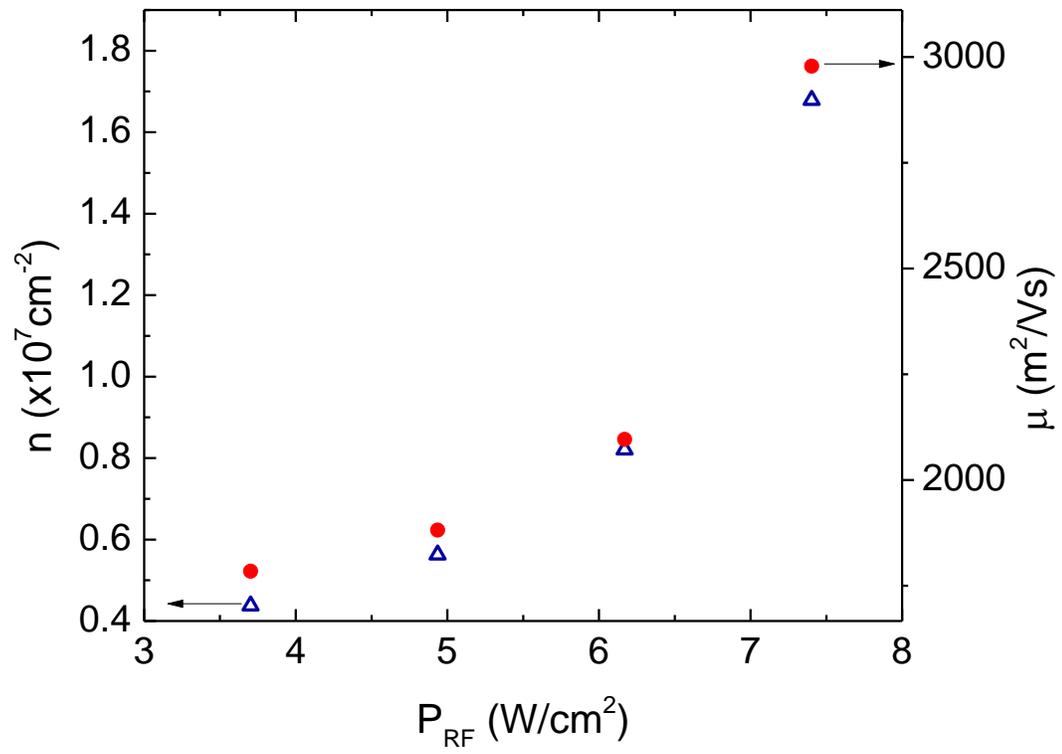

Fig. 7

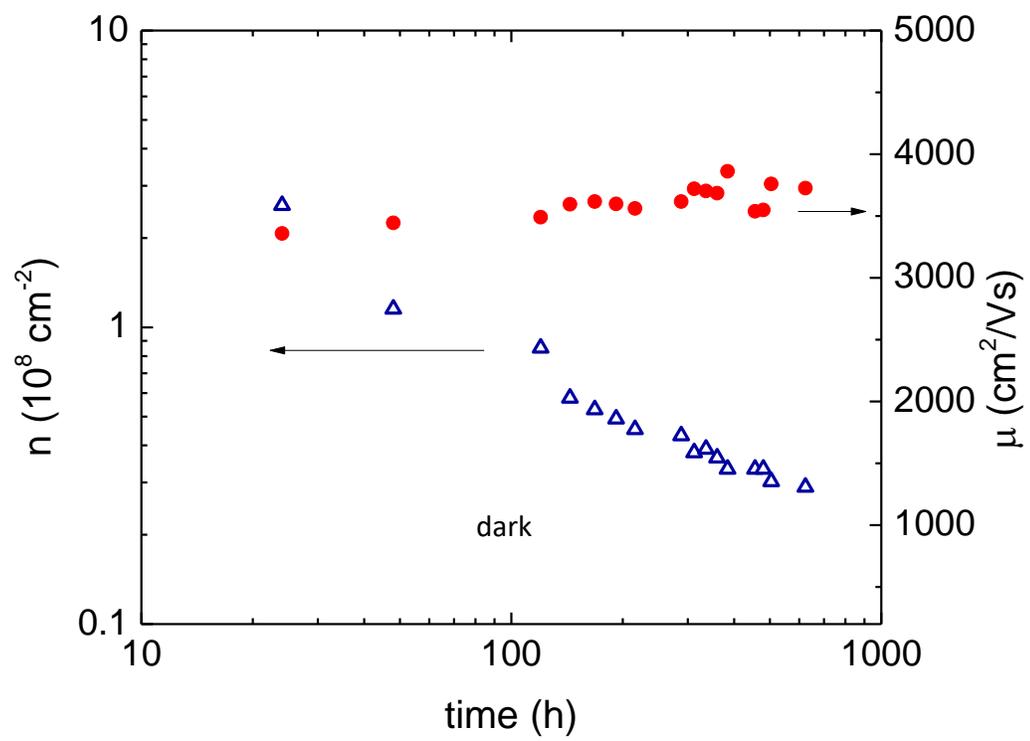

Fig. 8

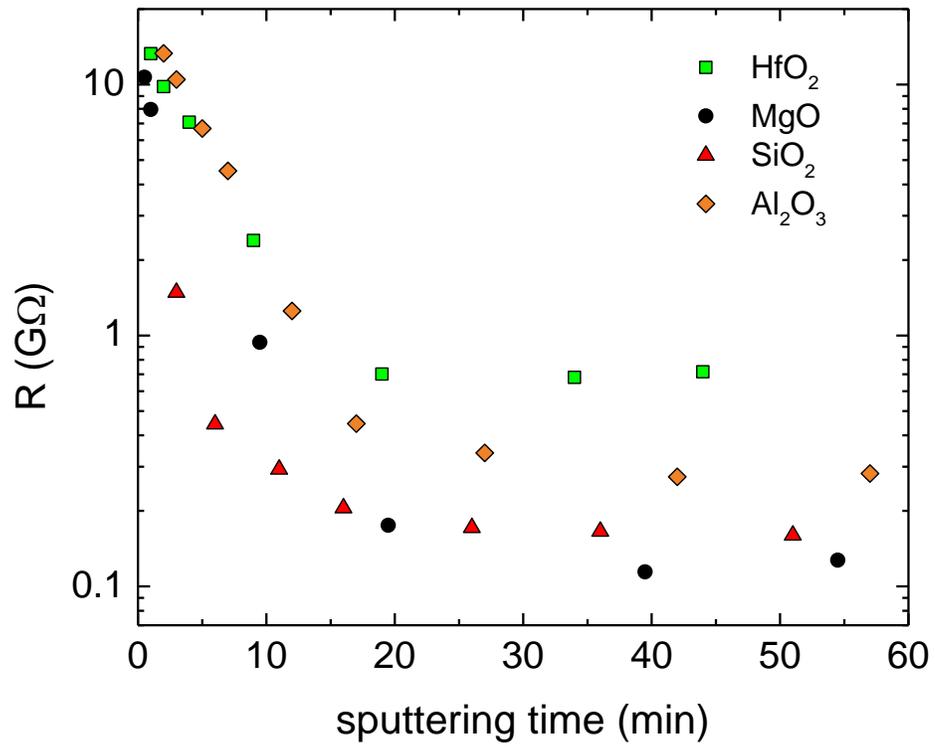

Fig. 9